\documentclass[ACS,STIX1COL]{WileyNJD-v2}

\articletype{Original article}%

\received{September 2023}
\revised{2023}
\accepted{2023}

\begin{document}

\title{Influence of the ion core on relaxation processes in dense plasmas}

\author[1,2]{T.S. Ramazanov}
\author[1,2]{S.K. Kodanova}
\author[1,2]{M.K. Issanova*}
\author[3]{B.Z. Kenzhegulov}
\authormark{Ramazanov T.S. \textsc{et al}}

\address[1]{\orgdiv{Institute of Applied Sciences and IT, 40-48 Shashkin str., 050038 Almaty, Kazakhstan}}

\address[2]{\orgdiv{Institute for Experimental and Theoretical Physics, Al-Farabi Kazakh National University, 71 Al-Farabi ave., 050040 Almaty, Kazakhstan}}

\address[3]{\orgdiv{Atyrau University named after K.Dosmukhamedov, 1 Studenchesky Ave., 060011 Atyrau, Kazakhstan}}
\corres{*M.K.~Issanova \email{issanova@physics.kz }}

\abstract{ The effect of an ionic core on the temperature relaxation in dense hot plasma of beryllium  is  studied using the pseudpotential model by Gericke \textit{et al} [Phys. Rev. E 2010, 81, 065401(R)]. Employing  the screened version of the ion pseudpotential [by Ramazanov \textit{et al}, Phys. Plasmas 2021, 28 (9), 092702], we computed the quantum transport cross-section for the electron-ion collisions in dense beryllium plasma, where screening is taking into account using the density response function in the long wavelength regime. The results for the transport cross-section are used to compute a generalised Coulomb logarithm and  electron-ion collision frequency. Utilizing the latter, we show the effect of the ionic core on the temperature relaxation. To understand the role of the ionic core, we compare the results with the data computed considering ions as point-like charges.
}


\jnlcitation{\cname{
\author{Ramazanov T.S.}, 
\author{Kodanova S.K.}, 
\author{Issanova M.K.}, 
\author{Kenzhegulov B.Z.},
\author{00000}} (\cyear{2023}), 
\ctitle{Influence of the ion core on relaxation processes in dense plasmas}, 
\cjournal{Contrib. to Plasma Phys.}, 
\cvol{2023;00:1--6}.}

\maketitle

\section{Introduction}\label{sec1}

Dense plasma is generated and studied in experiments due to the relevance for astrophysics and  inertial confinement fusion. 
In contrast to ordinary nearly ideal plasmas genereted at low densities, e.g., in gas discharges, the description of the dense plasmas requires taking into account electronic quantum degeneracy effect and correlation effects.
Furthermore, extreme conditions---due to high temperatures and densities---make the diagnostics of dense plasmas highly challenging \cite{Dornheim2022}. 
Therefore, theoretical studies and simulations are highly relevant for the understanding of the processes in dense plasmas \cite{pop_bonitz, Moldabekov2022_SciRep, jctc_2022}. 

The dense plasmas created in the laboratory experiments at such facilities as  the European XFEL \cite{Preston_2020} are often in a nonequilibrium  state. Particularly, the dense plasmas created by a shock compression and the dense plasmas generated by  laser radiation are well-known examples of nonequilibrium plasmas. In the first case, the shock energy is imparted mainly to the ions. The energy transfer from the ions to the electrons, which eventually brings the system to a thermodynamic equilibrium, takes place downstream of the shock over a comparatively long time \cite{Boerker,Hazak}. In the second example, laser energy couples mainly to the electrons. Also, in this case, equilibration between the electron and ion temperatures takes place over a time much longer than the thermalization within each subsystem.
Additionally, the non-isothermal plasmas also appear during the interaction of heavy-ion beams with a target.

The dense plasmas created by laser radiation  typically have the electrons that are hotter than the ions. \cite{Celliers, Ng}. Therefore, the knowledge of the energy transfer rate between electrons and ions is necessary to model and understand dense plasmas. Properties related to the energy transfer rate between electrons and ions are also needed to simulate the evolution of the dense plasmas generated in the inertial confinement fusion experiments. 

Recently, the effect of an ionic core (i.e, tightly bound states) on the electron-ion scattering in dense plasmas \cite{POP2021} and on the ionic transport characteristics  \cite{POP2022} has been investigated using a pseudopotential approach.
In this work, we extend these studies by analysing the the effect of an ionic core on the temperature relaxation  in dense plasmas.
We compute the Coulomb logarithm and electron-ion collision frequency using the method developed by  Rightley and Baalrud \cite{Rightley}, which is based on the quantum Boltzmann equation with Uehling-Uhlenbeck collision operator \cite{PhysRev.43.552}. The screening of the ion charge is taken into account using the density response function of the correlated electron gas in the long wavelength limit \cite{CPP_2022_Moldabekov, CPP_2017_Moldabekov}.

We aim to elucidate the change in temperature relaxation rates due to the effect of the ionic core in dense plasmas. 
For this, we consider hot dense beryllium and use the pseudopotential proposed by  Gericke \textit{et al.} \cite{Gericke1}.
To distinguish the effect of the ionic core on the temperature relaxation rates from other plasma effects such as screening, quantum degeneracy, and exchange-correlation effects, we 
compare with the results computed using the screened Thomas-Fermi (Debye) potential at the same plasma parameters.


The paper is organized as the following: In Sec.~\ref{sec2}, we describe the used screened ion-electron potential that takes into account the ionic core effect and screening due to electrons; In Sec. ~\ref{sec3}, the method used to compute the Coulomb logarithm, the electron-ion collision frequency, and the temperature relaxation time is presented; In Sec. ~\ref{sec4}, we present and discuss the results of the calculations. 

\section{Interaction potential}\label{sec2}

Gericke \textit{et al.} \cite{Gericke1} proposed a pseudopotential approach to describe the shielding of ion cores in plasmas, and found that the deviation of the pseudopotential from the Coulomb potential near the ion core is due to strongly bound electrons.
Following  Ref. \cite{Gericke1}, our analysis is based on the following general form of the pseudopotential:
\begin{equation} \label{potei} 
\varphi _{ei} (r)=\frac{Ze^{2} }{r} \left[1-\exp \left(-\frac{r^{\alpha } }{r_{\rm cut}^{\alpha } } \right)\right], 
\end{equation} 
where the parameter $\alpha$ controls the steepness of the core edge and,in combination with $r_{\rm cut}$, the depth of the minimum.

In the limit of  $\alpha \to \infty$, Eq.~(\ref{potei}) reduces to an empty core potential \cite{Ashcroft1, Ashcroft2}. 
The soft form of potential (\ref{potei}) helps to avoid apperance of such nonphysical effects as an infinite force at the core edge.
Furthermore, two parameters $r_{\rm cut}$ and $\alpha$  together allow to fit pseudo-potentials developed for Kohn-Sham density functional simulation (KS-DFT), which is proved to provide an accurate description of dense plasmas in the case of a partial degeneracy.
Additionally, at $\alpha=1$,  Eq.~(\ref{potei}) reduces to the well known Deutsch potential \cite{Deutsch1} which was successfully used for the calculation of various properties of dense hydrogen plasmas with weak electronic degeneracy \cite{Deutsch2, Kodanova1,Kodanova2, Moldabekov1, MRE18, Moldabekov_cpp_12, issanova_2016, gab_16}.  

Potential ~(\ref{potei}) takes into account the ion core effect, but does not have screening due to plasma electrons.
We are going to use the electron-ion interaction potential for the calculation of the electron-ion scattering cross section.  The latter we use to compute the relaxation rate of the electron-ion temperature in the pair-collision approximation. Therefore, we need the screened version of potential ~(\ref{potei}), which reads \cite{POP2021}:
\begin{equation}\label{potIon}
\Phi_{ei} =\frac{Ze}{r} \exp(-k_{s} r )\left[1-\exp \left(-\frac{r^{\alpha } }{r_{\rm cut}^{\alpha } } \right)\right],
\end{equation}
where the inverse screening length $k_s=1/\lambda_s$ is defined as: 
\begin{equation}\label{KS}
k_{s}^{2} =k_{id}^{2} /(1-k_{id}^{2} \gamma ),\quad \quad k_{id}^{2} =k_{TF}^{2} \theta ^{{1\mathord{\left/ {\vphantom {1 2}} \right. \kern-\nulldelimiterspace} 2} } I_{{-1\mathord{\left/ {\vphantom {-1 2}} \right. \kern-\nulldelimiterspace} 2} } (\eta )/2,
\end{equation}
where $\eta =\mu /k_{B} T_{e} $ is the reduced chemical potential, $I_{-{1\mathord{\left/ {\vphantom {1 2}} \right. \kern-\nulldelimiterspace} 2} } $ the Fermi integral of the order $-1/2$ and $k_{TF} =\sqrt{3} \omega_{p} /\upsilon_{F}$  the Thomas-Fermi wavenumber,  $\omega_{p} $ the plasma frequency, and $\upsilon_{F}=k_F\hbar/m$ is the Fermi velocity. 
The parameter $\gamma$ takes into account the local field correction (exchange-correlation effect) in the long wave-length limit \cite{CPP_2022_Moldabekov}:
\begin{equation}
    \gamma=-\frac{k_F^2}{4\pi e^2}\frac{\partial^2{nf_{\rm xc}(n,T_e)}}{\partial n^2},
\end{equation}
with $f_{\rm xc}(n,T_e)$ being the electronic exchange-correlation free energy density of the uniform electron gas (UEG) at density $n$ and temperature $T_e$.
To compute $f_{\rm xc}(n,T_e)$, we use  the quantum Monte-Carlo simulations-based parameterization by Groth et al. \cite{PhysRevLett.119.135001}

Screened potential (\ref{potIon}) was derived using potential (\ref{potei}) and the density response function of UEG in the long-wavelength approximation \cite{pop_qhd_2018, pop_ion_bohm_2015, cpp_bohm_2017}. If one neglects the ion core effect, screened potential (\ref{potIon}) reduces to a standard screened potential of a test charge, which is often referred to as Yukawa potential. 
To analyse the effect of the ionic core on the temperature relaxation in dense plasmas, we compare the the results computed using potential (\ref{potIon}) with the data calculated using screened Coulomb potential:
\begin{equation}\label{Yukawa}
\Phi_{Y} =\frac{Ze}{r} \exp(-k_{s} r ),
\end{equation}
which is referred to as the \textit{Yukawa potential}. 

Before further discussion of Eq.~(\ref{potIon}), we need first introduce the dimensionless plasma parameters used in this work.
There are two dimensionless parameters characterizing the state of the system:  the density parameter $r_{s} =a/a_{B} $ defined as the ratio of the mean inter-electronic distance  $a=\left(\frac{3}{4\pi n} \right)^{1/3} $ to the first Bohr radius $a_{B} =\hbar ^{2} /\left(m_{e} e^{2} \right)$ and the degeneracy parameter $\theta =\frac{k_{B} T}{E_{F} }$ defined as the ratio of the thermal energy $k_B T$ to the Fermi energy  $E_{F}$. 

\begin{figure}[t]
\includegraphics[width=1\textwidth]{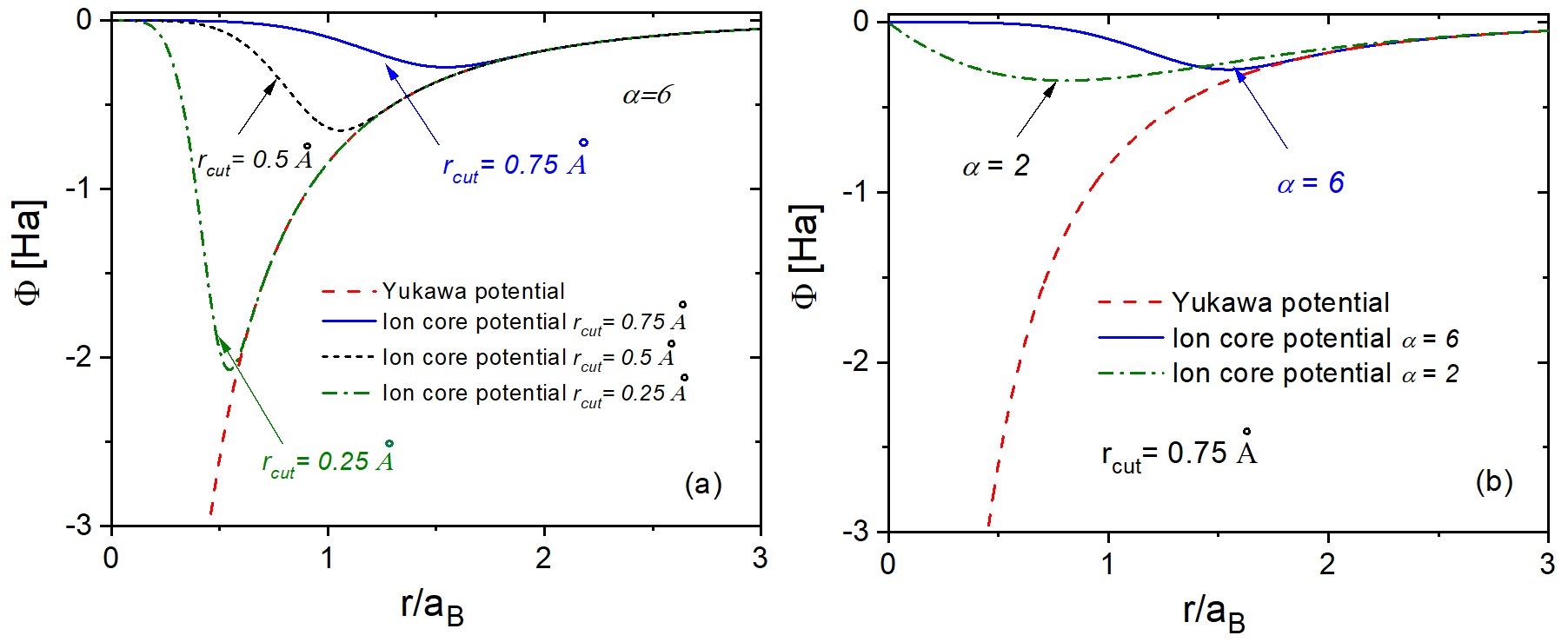}
\caption{(a) Electron-ion interaction potential (\ref{potIon}) with $\alpha=6$ and the Yukawa potential (dashed red line) at $\theta=1$ and $r_{S}=2$ for different values of $r_{\rm cut}$, (b) Electron-ion interaction potential (\ref{potIon}) with $r_{\rm cut}=0.75$ and the Yukawa potential (dashed red line) at $\theta=1$ and $r_{S}=2$ for different values of $\alpha$.}
 \label{fig:1}
\end{figure}

Potential (\ref{potIon}) agrees with the screened Coulomb potential~(\ref{Yukawa}) (Yukawa potential) at large distances and  reduces to the bare  pseudopotential (\ref{potei}) close to the ionic core.
In this work we consider hot dense beryllium. Following  the analysis of the pseudopotential of the beryllium ion by Gericke \textit{et al.} \cite{Gericke1}, we consider $\alpha=2$ and $\alpha=6$, and set the cutoff radius for the beryllium ion to $r_{\rm cut} =0.75{\kern 1pt} \, {\mathop{{\rm A} }\limits^{\circ }}\simeq 1.42~{a_B}$. This value of $r_{\rm cut}$ is found by fitting to the  pseudopotential generated for the applications within Kohn-Sham density functional theory (KS-DFT).

A detailed discussion of potential (\ref{potIon}) is provided in Ref. \cite{POP2021}. Here, for the illustration of the general features of  potential (\ref{potIon}),  we show potential (\ref{potIon}) in Fig. \ref{fig:1} at $\theta=1$ and $r_s=2$. In Fig. \ref{fig:1} (a), we set $\alpha=6$ and show results for different values of $r_{\rm cut}$. Additionally, screened potential (\ref{potIon}) is compared with the Yukawa potential. At fixed $\alpha$, the increase in $r_{\rm cut}$ results  in a softer potential at small distances and the decrease in $r_{\rm cut}$ results in a deeper negative minimum of the potential. In Fig. \ref{fig:1} (b), we set $r_{\rm cut}=0.75$  and compare screened potential values at $\alpha=6$ and $\alpha=2$. The decrease in $\alpha$ leads to closer values of the distance from an ion at which   potential (\ref{potIon}) has its minimum. We note that at considered parameters, we see the deviation of  potential (\ref{potIon}) from the Yukawa potential at $r\lesssim 2 a_B$. Also, we note that the deviation of  potential (\ref{potIon}) (at $\alpha=6$ and $\alpha=2$) from the Yukawa potential is much larger than the difference of the values of  potential (\ref{potIon})  at $\alpha=6$  from that at $\alpha=2$.

\section{Relaxation characteristics}\label{sec3}

The relaxation rates of the electron and ion temperatures are determined by the  collision rates (frequencies) between electrons and ions and the difference of  electron and ion temperatures  ~\cite{Glosli}:
\begin{equation} \label{dT}
\frac{dT_{e} }{dt} =\frac{T_{i} -T_{e}}{\tau_{ei}},\,\,
\frac{dT_{i} }{dt} =\frac{T_{e} -T_{i}}{\tau_{ie}},
\end{equation}
where the collision rates read:
\begin{equation} \label{nu}
\nu_{ei} = \frac{1}{\tau_{ei}} = \frac{8\sqrt{2\pi} n_{i}Z^{2}e^{4} \Xi}{3 m_{e} m_{i}} \left(\frac{k_{B}T_{e}}{m_{e}}+\frac{k_{B}T_{i}}{m_{i}}\right)^{-3/2}\,\text{and}\,\, \nu_{ie} = \frac{1}{\tau_{ie}}= Z \nu_{ei},
\end{equation}
with $\Xi$ being a generalized Coulomb logarithm \cite{Rightley}, which takes into account  both the effects of degeneracy and strong coupling.

The generalized Coulomb logarithm is compute using the transport cross-section $Q^{T}$ defined by electron-ion collisions:
\begin{equation} \label {Xi}
 \Xi= {\frac{1}{2} \int_{0}^{\infty} dg G(g) \frac{Q^{T}(g)}{\sigma_{0}}},
\end{equation}
where $\sigma_{0} =\hbar ^{2} /(m_{e} v_{Te})^{2}$, $g=u/v_{Te}$, and $u$ is the relative velocity of the scattering particles. 

In Eq. (\ref{Xi}), the function $G(g)$ takes into account the Fermi-Dirac statistics of electrons \cite{Rightley}:

\begin{equation} \label {G}
G(g)= {\frac{\eta \exp^{-g^{2}}g^{5}} {[-Li_{3/2}(-\eta)](\eta e^{-g^{2}}+1)^{2}}}.
\end{equation}
where $-{\rm Li}_{3/2}=[-\eta]=\frac{4}{3\sqrt{\pi}} \theta^{-3/2}$, $\eta=\exp(\mu/k_{B}T_e)$, and $\mu$ is the electron chemical potential \cite{Melrose, Rightley}.

The  transport cross-section $Q^{T}$ is computed according to two-particle quantum scattering theory \cite{Massey1965}:
\begin{equation}\label{Transport}
Q^{T} (k)=\frac{4\pi }{k^{2} } \sum _{l}(l+1)\sin ^{2}  (\delta _{l} (k)-\delta _{l+1} (k)),
\end{equation} 
where the partial cross-section $Q_{l}$ reads
\begin{equation}
\label{Elastic} 
Q_{l} (k)=\frac{4\pi }{k^{2} } (2l+l)\sin ^{2} \delta _{l} (k),
\end{equation} 
and the phase shift  $\delta _{l} (k)\equiv \delta _{l} (k,r\to \infty )$ is calculated by solving the Calogero equation ~\cite{Calogero, Babikov, Babikov0}:
\begin{equation}\label{Phase} 
    \frac{d\delta _{l} (kr)}{dr} =-\frac{1}{k} \Phi (r)\left[\cos \delta _{l} (kr)j_{l} (k,r)\right.-\left.\sin \delta _{l} (k,r)n_{l} (k,r)\right]^{2},
\end{equation} 
with the condition $\delta _{l} (k,0)=0$.
In Eq.~(\ref{Phase}), $k=mu/\hbar$ is the wave number, $l$ indicates the orbital quantum number, $j_{l}$ and $n_{l} $ are the Rikkati-Bessel functions, and $\Phi(r)$ is the pair interaction potential between colliding particles.
In this work, we consider potentials (\ref{potIon}) and (\ref{Yukawa}).

\section{Results and discussion}\label{sec4}

\begin{figure}[t]
\centerline{\includegraphics[width=480pt,height=16pc]{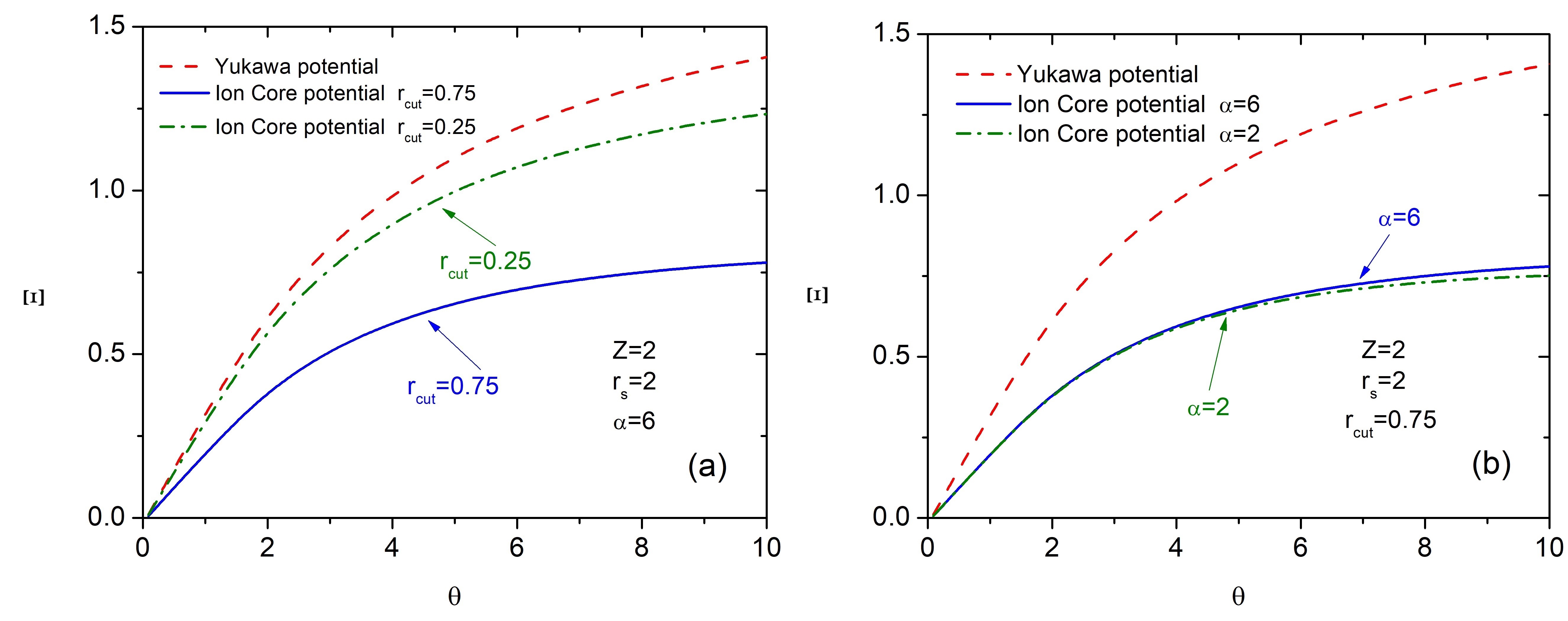}}
\caption{The Coulomb logarithm at $\theta=1$ and $r_s=2$. The left panel (a) is for $\alpha=6$ with  $r_{\rm cut}=0.25$ and $r_{\rm cut}=0.75$. The right panel (b) is for $r_{\rm cut}=0.75$ with $\alpha=2$ and $\alpha=6$.}
\label{fig:3}
\end{figure}

Let us start the discussion from the changes in the Coulomb logarithm $\Xi$ due to the ionic core effect.
Fig. \ref{fig:3} shows the Coulomb logarithm $\Xi$  as a function of $\theta$  in the case (a) $\alpha=6$ for different values of $r_{\rm cut}$ ($r_{\rm cut}=0.25$ to $r_{\rm cut}=0.75$ ) and in the case (b) $r_{\rm cut}=0.75$ for different values of $\alpha$. In Fig. \ref{fig:3} (a), the red dashed line indicates the results computed using the Yukawa potential, the blue solid line corresponds to the results computed using $r_{\rm cut}=0.75$, and the green dash-dotted line is the results computed by setting $r_{\rm cut}=0.25$. From  Fig. \ref{fig:3} (a), we see that the use of the pseudopotential (\ref{potIon}) results in smaller values of the  Coulomb logarithm with the increase in $r_{\rm cut}$. This is expected since the increase in the parameter $r_{\rm cut}$ leads to the weaker electron-ion interaction (as one can see from Fig. \ref{fig:1}).  

\begin{figure}[t]
\centerline{\includegraphics[width=480pt,height=16pc]{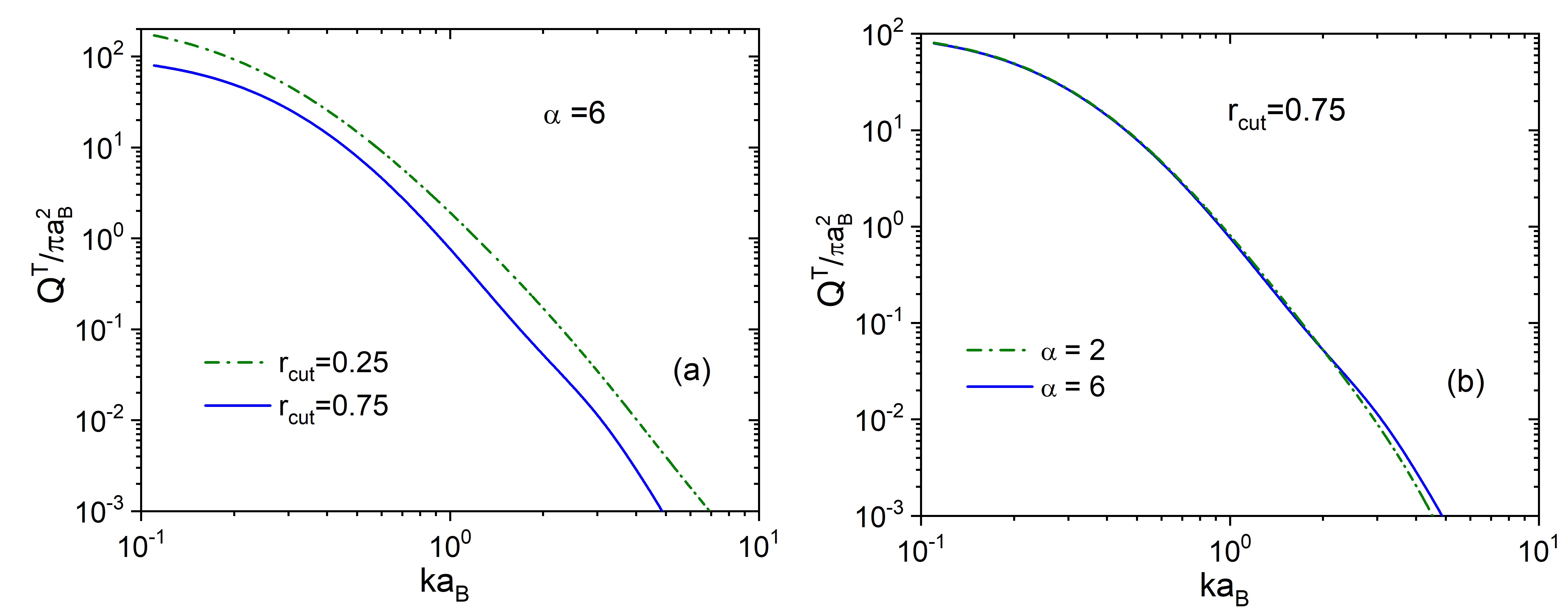}}
\caption{The transport cross-sections at $r_s=2$ and $\theta=1$.The left panel (a) is for $\alpha=6$ with  $r_{\rm cut}=0.25$ and $r_{\rm cut}=0.75$. The right panel (b) is for $r_{\rm cut}=0.75$ with $\alpha=2$ and $\alpha=6$.}
\label{fig:2}
\end{figure}

In Fig. \ref{fig:3} (b), we show the Coulomb logarithm $\Xi$ at $r_{\rm cut}=0.75$ for $\alpha=2$ and for $\alpha=6$. We compare the results with the data computed using the Yukawa potential.
From this figure we see that the change from $\alpha=2$ to $\alpha=6$ does not create significant changes in the Coulomb logarithm. To understand that, we recall that the change  of $\alpha$ from $\alpha=2$ to $\alpha=6$ does not lead to substantial  changes in the potential. To quantify the effect of the change of $\alpha$ and $r_{\rm cut}$ for electron-ion scattering, we show in Fig. \ref{fig:2}   the transport cross-sections computed at $r_s=2$ and $\theta=1$  for (a) $\alpha=6$ with different values of $r_{\rm cut}$ and for (b) $r_{\rm cut}=0.75$ with  $\alpha=2$ and $\alpha=6$ . 
From Fig. \ref{fig:2}  (a), we see that the increase in $r_{\rm cut}$ from $r_{\rm cut}=0.25$ to $r_{\rm cut}=0.75$ significantly decreases the transport cross section. This results in the decrease in the Coulomb logarithm $\Xi$. In contrast, from Fig. \ref{fig:2}  (b), we see that the change in  $\alpha$ from $\alpha=2$ to $\alpha=6$  at fixed $r_{\rm cut}=0.75$  does not affect much the the transport cross section and as the result the Coulomb logarithm  does not change much as well. We note that a detailed study of the electron-ion quantum scattering cross sections using potential (\ref{potIon}) was performed in Ref. \cite{POP2021}.

\begin{figure}[t]
\centerline{\includegraphics[width=480pt,height=16pc]{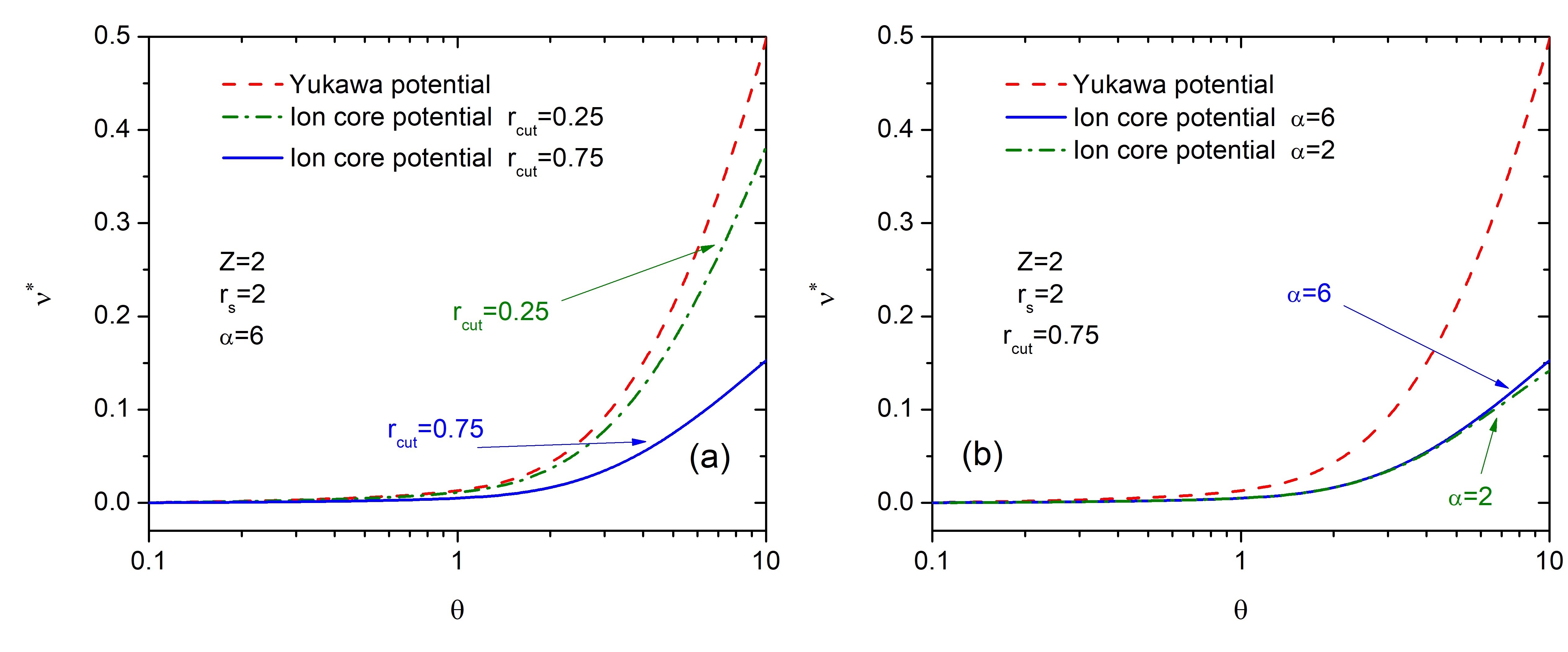}}
\caption{The frequency of the electron scattering in the field of potential (\ref{potIon}) at $r_s=2$ and different values of $\theta$ (in units of plasma frequency). The left panel (a) is for $\alpha=6$ with  $r_{\rm cut}=0.25$ and $r_{\rm cut}=0.75$. The right panel (b) is for $r_{\rm cut}=0.75$ with $\alpha=2$ and $\alpha=6$. }
\label{fig:4}
\end{figure}

Next, according to Eq. (\ref{nu}), we compute the electron-ion collision frequencies using the the Coulomb logarithm $\Xi$ defined in  Eq. (\ref{Xi}). Fig. \ref{fig:4} shows the results for the electron-ion collision frequency  as a function of $\theta$  in the case (a) $\alpha=6$  with  $r_{\rm cut}=0.25$ and $r_{\rm cut}=0.75$ and in the case (b) $r_{\rm cut}=0.75$ with $\alpha=2$ and $\alpha=6$. The results are computed using potential (\ref{potIon}) and the Yukawa potential. In agreement with the behavior of the transport cross section and the Coulomb logarithm, the increase in $r_{\rm cut}$ from $r_{\rm cut}=0.25$ to $r_{\rm cut}=0.75$ at fixed $\alpha=6$ value results in the substantial decrease in the electron-ion collision frequency.
In contrast, the change in $\alpha$ from $\alpha=2$ to $\alpha=6$ at fixed $r_{\rm cut}=0.75$  almost does not affect the electron-ion collision frequency. 
In general, at considered parameters, the electron-ion collision frequency values computed using potential  (\ref{potIon}) are smaller than the electron-ion collision frequency values obtained using the Yukawa potential. 

\begin{figure}[t]
\centerline{\includegraphics[width=480pt,height=16pc]{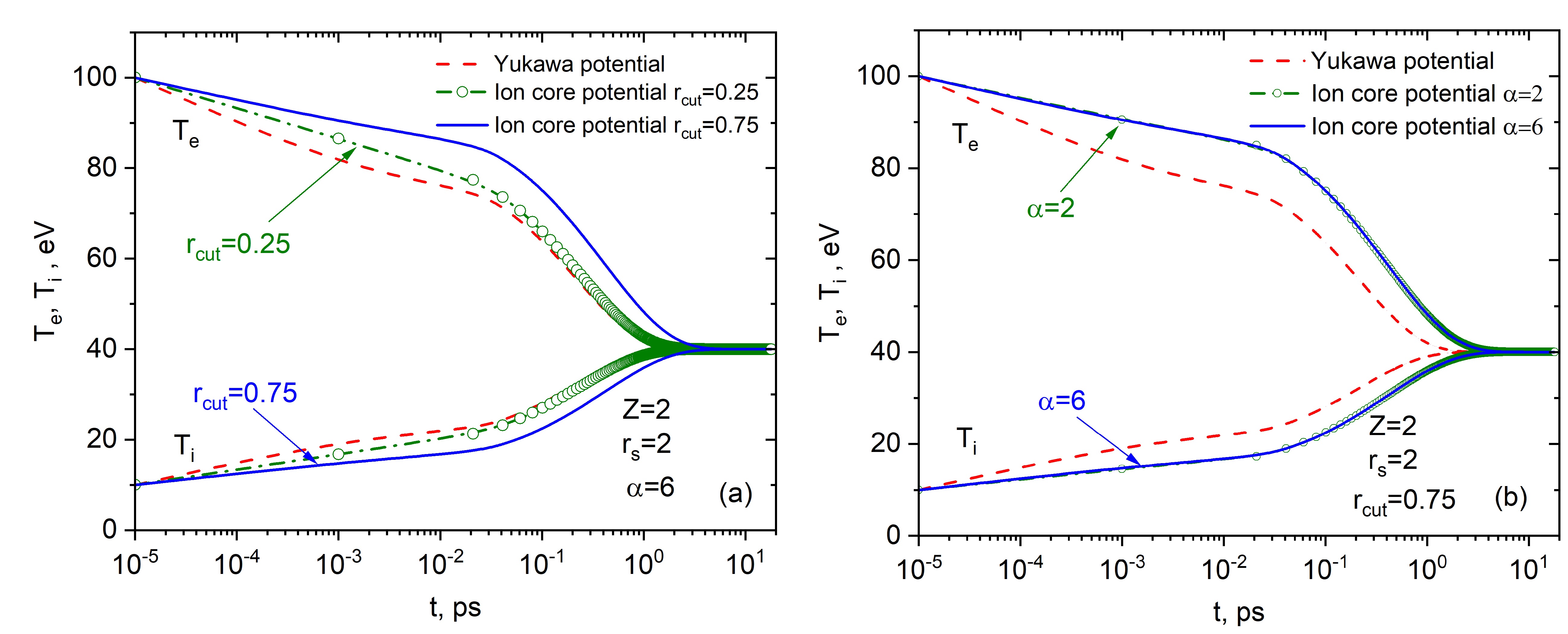}}
\caption{The relaxation of temperature between electrons and ions in dense beryllium plasmas at $r_s=2$. The left panel (a) is for $\alpha=6$ with  $r_{\rm cut}=0.25$ and $r_{\rm cut}=0.75$. The right panel (b) is for $r_{\rm cut}=0.75$ with $\alpha=2$ and $\alpha=6$.}
\label{fig:5}
\end{figure}

Now, to demonstrate the effect of ionic core on the temperature relaxation in beryllium plasma with electrons hotter than ions,  we solve Eqs. (\ref{dT})-(\ref{Phase}) self-consistently. 
Our results for the relaxation of temperatures of electrons and ions are presented in Fig. \ref{fig:5}, where electrons have the initial temperature of $100~{\rm eV}$ and ions have the initial temperature $10~{\rm eV}$.  In Fig. \ref{fig:5}, we consider (a) $\alpha=6$ with different values of $r_{\rm cut}$ and (b) $r_{\rm cut}=0.75$ with different values of $\alpha$.  
From Fig. \ref{fig:5}, we see that the ion core effect lead to a slower relaxation of the temperatures to an equilibrium value compared to the results computed using the Yukawa potnetial.
From Fig. \ref{fig:5}(a), it is clear that the decrease in $r_{\rm cut}$ from $r_{\rm cut}=0.75$ to $r_{\rm cut}=0.25$ results in faster temperature relaxation. This is expected since the collision frequency increases with the decrease in $r_{\rm cut}$. On the other hand, one can say that the reduction in $r_{\rm cut}$ leads to the better agreement with the model where ions considered to be point charges. From Fig. \ref{fig:5}(b), we clearly observe that at the fixed value $r_{\rm cut}=0.75$, the increase of the parameter $\alpha$ from $\alpha=2$ to $\alpha=6$ does not affect the temperature relaxation time.


\begin{figure}[t]
\centerline{\includegraphics[width=480pt,height=16pc]{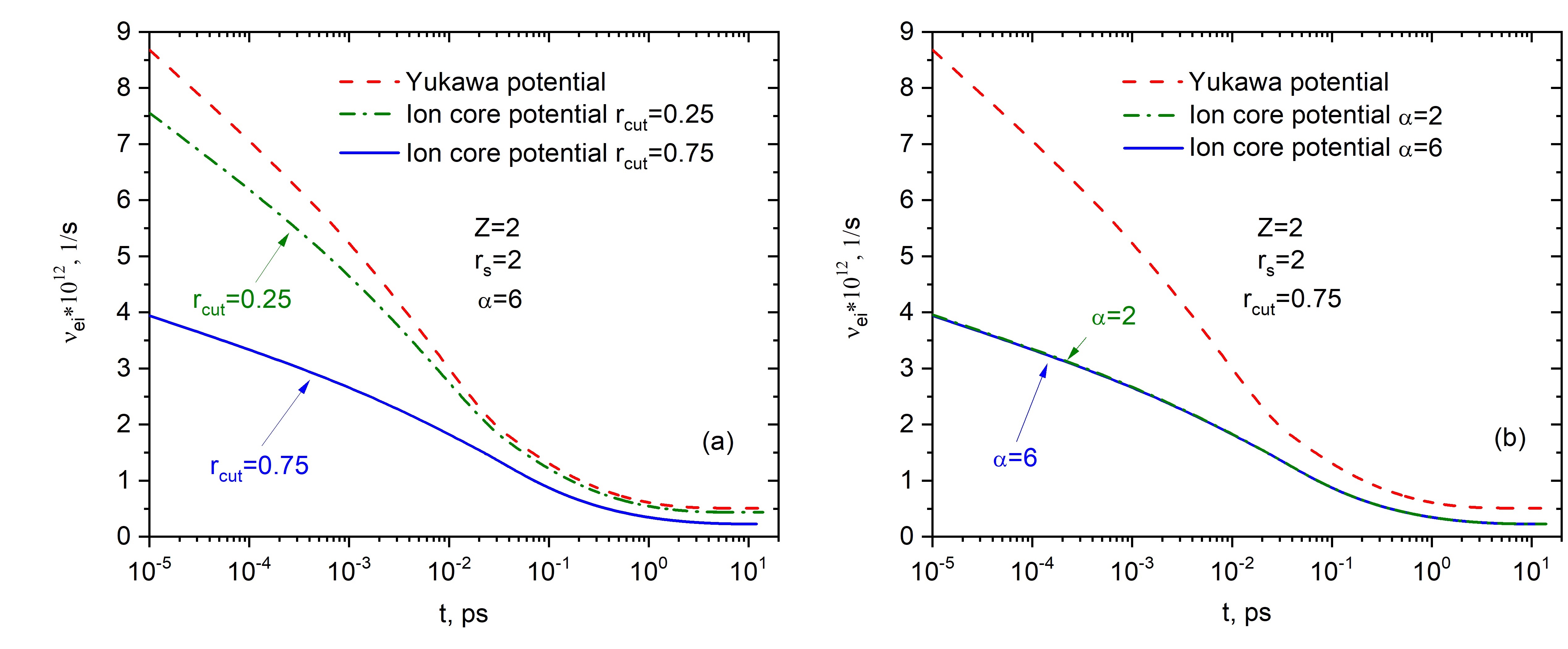}}
\caption{The electron-ion collision frequency during the temperature relaxation process in dense beryllium plasmas at $r_s=2$. The left panel (a) is for $\alpha=6$ with  $r_{\rm cut}=0.25$ and $r_{\rm cut}=0.75$. The right panel (b) is for $r_{\rm cut}=0.75$ with $\alpha=2$ and $\alpha=6$.}
\label{fig:6}
\end{figure}

Finally, according to temperature relaxation calculations presented in Fig. \ref{fig:5}, we show in Fig. \ref{fig:6} the  self-consistently computed change in the electron-ion collision frequency during the temperature relaxation period. 
In Fig. \ref{fig:6}, the left panel (a) shows the results for  $\alpha=6$ with different values of $r_{\rm cut}$ and the right panel (b) presents the data for $r_{\rm cut}=0.75$ with different values of $\alpha$. From Fig. \ref{fig:6}, we see that the electron-ion collision frequency  reduces drastically with approaching equilibrium state. The Yukawa model for the potential of ions leads to significantly larger values of the  the electron-ion collision frequency during the entire relaxation process. From Fig. \ref{fig:6} (a) one can observe that the increase in $r_{\rm cut}$ is accompanied by the decrease  in the electron-ion collision frequency  and from Fig. \ref{fig:6} (b) we conclude that we have nearly the same value of the electron-ion collision frequency for  $\alpha=2$ and $\alpha=6$ at $r_{\rm cut}=0.75$.


\section{Conclusions}\label{last}

The effect of the ionic core on the temperature relaxation was analyzed considering hot dense beryllium plasma. 
The screened potential of the electron-ion interaction was used to compute the quantum scattering transport cross-section for the electron-ion collisions in dense plasmas. 
The ion-core effect is taken into account using the pseudopotential approach suggested by Gericke et al \cite{Gericke1}. The effect of screening is taken into account using the long-wavelength approaximation for the desnity response of UEG \cite{POP2021}.
The data for transport cross-section was utilized to compute the generalized Coulomb logarithm and the temperature relaxation in hot dense plasma with the initial temperature of electrons $100~{\rm eV}$ and with colder ions with the inital temperature of $10~{\rm eV}$. The results of the calculations clearly show that the ion core effect leads to a slower temperature relaxation compared to the case where ions treated as point-like charges.  


\section*{Acknowledgments}
This research was supported by the Ministry of Science and Higher Education of Kazakhstan, under Grant AP19677200 "Investigation of structural and dynamic properties of non-ideal plasma".

\bibliography{wileyNJD-ACS}%

\end{document}